\documentclass[a4paper,11pt]{article} 
\usepackage{pos}
\usepackage{pgf}
\usepackage{wrapfig}
\usepackage[british]{babel}

\newcommand{\vevT}[1]{\langle 0 \left| T\left\{ #1 \right\} \right| 0\rangle}

\newcommand{\GeV}{\mathrm{GeV}}
\newcommand{\Tr}{{\mathrm{Tr}~}}
\title{A review on Glueball hunting} \ShortTitle{Review on Glueball Hunting}

\author*[a]{Davide Vadacchino}

\affiliation[a]{Centre for Mathematical Sciences, University of Plymouth,
Plymouth, PL4 8AA, United Kingdom}

\emailAdd{davide.vadacchino@plymouth.ac.uk}

\abstract{One of the most direct predictions of QCD is the existence of
color-singlet states called Glueballs, which emerge as a consequence of
the gluon field self-interactions.

Despite the outstanding success of QCD as a theory of the strong interaction
and decades of experimental and theoretical efforts, 
all but the most basic
properties of Glueballs are still being debated.

In this talk, I will review efforts aimed to understanding Glueballs and the
current status of Glueball searches, including recent experimental results and
lattice calculations.  }

\FullConference{The 39th International Symposium on Lattice Field Theory,\\
8th-13th August, 2022,\\ Rheinische Friedrich-Wilhelms-Universität Bonn, Bonn,
Germany }

\begin{document} \maketitle
	
\section{Introduction} 

Quantum Chromodynamics (QCD) is believed to be the microscopic theory of the 
strong interaction. It has been very successful at explaining a wide range of
experimental results, especially in the high-energy regime, where
perturbation theory is applicable. As the target energy scale is lowered, 
and the coupling grows, perturbation theory is not viable anymore.
The relationship between the degrees of freedom present in the Lagrangian
density and the phenomenology becomes opaque.
Despite the lack of a proof from first principles, confinement allows to
restrict the realized states to the color singlets. Hence, not only are
meson and baryon states predicted, but a plethora of states for which solid evidence
has begun to surface only in the last few years. Ironically, it is one of the earliest
predicted states, the \emph{glueball}, that still await undisputed experimental 
confirmation.

Glueballs are quarkless color-singlet states of QCD. Their hypothetical spectrum 
and decay patterns have been the object of studies for more than $50$ years, leaving
an important footprint in the literature. Glueballs have proven to be 
very elusive objects: despite eclectic approaches, 
the community only agrees on their basic properties and an understanding of
the link between their macroscopic properties and the underlying 
Yang-Mills dynamics is still missing. Yet, they are one of the most distinctive
predictions of QCD and essential to the confirmation of every aspect of the theory.

In this talk, past and present efforts to determine the spectrum and decay
patterns of Glueballs, an activity called ``glueball hunting'', will be reviewed.
Phenomenological approaches, based on the intuition gained from the quark model, 
were historically the first to be developped and are reviewed in 
Section~\ref{sec:pheno}. 
More recent analytical approaches, deeply rooted in QCD and based on the wealth
of results gained from modern computational techniques are the  focus of 
Section~\ref{sec:analytic}. Lattice calculations, the only first principles fully
non-perturbative approach capable of providing ready-for-comparison numbers, are 
reviewed in Section~\ref{sec:lat}. At last, a sample of the approach
to the experimental identification of a glueball is described in Section~\ref{sec:exp}.

\section{Phenomenological approaches} \label{sec:pheno}

The first reference to a glueball is found in 
Ref.~\cite{Fritzsch:1972jv}, where it is described 
as a state generated by a  
local color-singlet product of gluon fields $G_{\mu\nu}^a$ with 
isospin and $G$-parity $I^G=0^+$. The success of
the minimal quark model suggests that we proceed
by analogy, and build glueballs by progressively adding
gluons to the system while ensuring symmetry 
under the interchange of its constituent gluons. 
The full classification can be found
in Ref.~\cite{Jaffe:1985qp}. The simplest states
are obtained for $2$ and $3$ gluons. For $2$ gluons,
the only color-singlet operator is $\Tr G_{\mu\nu} 
G_{\rho\sigma}$, where the trace is on the color indices that
have been omitted. The decomposition in irreducible representations
of the Lorentz group result in,
\begin{equation}
\Tr G_{\mu\nu} G^{\mu\nu}~,\quad
\Tr \tilde{G}_{\mu\nu} G^{\mu\nu}~,\quad
\Tr G_{\alpha\nu} G^{\nu}_\beta - 
\frac{1}{2}g_{\alpha\beta} \Tr G_{\mu\nu} G^{\mu\nu}~,
\end{equation}
where $\tilde{G}_{\mu\nu}=\tfrac{1}{2}\epsilon_{\mu\nu\rho\sigma}G^{\rho\sigma}$
and $g$ is the metric tensor. For $3$ gluons, there are two 
color-singlet combinations,
\begin{equation}
    f_{abc} G_a^{\mu\nu} G_b^{\alpha\beta} G_c^{\delta\sigma},\quad
    d_{abc} G_a^{\mu\nu} G_b^{\alpha\beta} G_c^{\delta\sigma}~,
\end{equation}
where $f_{abc}$ are the structure constant of $SU(3)$ and $d_{abc}$
the related totally symmetric tensor. 

Since a single gluon has $j^\pi=1^-$, the enumeration is 
the following,
\begin{equation}
    J^{PC}=\begin{cases}
        (even\geq0)^{\pm+}\\
        (odd\geq3)^{++}
            \end{cases}
    ,\quad
    J^{PC}=\begin{cases}
        (odd\geq1)^{\pm+}\\
        (odd\geq3)^{--}
            \end{cases}~,
\end{equation}
for $2$ and for $3$ gluon states, respectively. 
Note that if the gluons are thought of as non-interacting and 
on-shell, the classification of possible states is analogous to the 
classification of two-photon states,
see Ref.~\cite{Landau:1948kw,Yang:1950rg}, and no $1^{-+}$ appears
among the $2$ gluon states.

Following Ref.~\cite{Jaffe:1985qp}, we can obtain an heuristic picture
of the spectrum by assuming that the mass of each states is 
proportional to the dimension of the operator that creates it. 
Hence, the lightest states are $0^{++}$, $0^{-+}$ and $2^{++}$,
known as scalar, pseudo-scalar and tensor glueballs, while
glueball with exotic $J^{--}$ quantum numbers will be found at
higher energies. There is no a priori reason to expect these
states to be stable. In the limit in which flavor-breaking effects
can be neglected, the decay widths $\Gamma(J^{PC})$ are flavor
agnostic and reproduce the branching ratios expected by $SU(3)$ 
symmetry. 
Moreover, they obviously are not the only states with 
$I^G=0^+$; as a consequence, nothing prevents them from mixing
with $q\bar{q}$ states of a similar mass if there are any. 
Below, we will focus on 
the $0^{++}$, $0^{-+}$ and $2^{++}$ states only, as they are 
plausibly the most accessible experimentally.

In the simplest approach, a glueball is a bound states of gluons
of constituent mass $\mu$, interacting through a potential. 
In Ref.~\cite{Barnes:1981ac}, a (massless) one-gluon-exchange potential is
obtained from the $O(g^2)$ two-gluon scattering diagrams prescribed
by QCD, supplemented by a linear confining 
potential with slope $\sigma_a$, the (adjoint) string tension. Clearly
the former is believed to describe short-range interactions, while
the latter the long-range confinement property.
The relative positions and splitting of $2$-gluons states
are then calculated in units of $\mu$. The, $(2n)^{\pm+}$ states 
are found to be degenerate at $O(g^2)$ level, and ordered 
according to the value of $n$.
Hence, the lightest states are $0^{\pm +}$, followed
by $2^{++}$. In particular, $m(0^{\pm+})=2.180\cdot 2\mu$ for the
ground state, followed by $m(0^{++,\star})/m(0^{\pm+})\simeq 1.4$. 
and $m(2^{++})/m(0^{\pm+})\simeq 1.16$, where the $\star$ indicates an
excited state. The value of $\mu$ is discussed and it is recognized
that when $\mu\to 0$, the only remaining scale is 
$\sqrt{\sigma_a}$ which is, unfortunately, inaccessible. Setting, 
in alternative $\sigma_a\simeq\sigma=0.4\,\GeV$, where $\sigma$ is
the fundamental string tension, then $m(0^{\pm+}) \simeq 1.5\,\GeV$.
As a result, $m(2^{++})=1.74\,\GeV$.

In Ref.~\cite{Cornwall:1981zr}, a more sofisticated attempt is done
at defining a constituent gluon mass.
Na\"ively one would like to define a constituent mass as the pole
mass of the gluon propagator. However, the latter is not physical,
being gauge variant. However, a rearrangement of the Feynman 
diagrams contributing to it can be defined, that satisfies the 
Slavnov-Taylor identities of gauge invariance. 
See Ref.~\cite{Aguilar:2006gr} for a recent discussion. The constituent gluon 
propgator $d(q^2)$ is defined by
\begin{equation}
    d^{-1}(q^2)=
    \beta_0 g^2 (q^2-\mu^2)\ln{\left[(4\mu^2-q^2)\Lambda^{-2} \right]}
    ,\quad
    \mu^2(q^2)= \mu^2 \left( 
        \frac{\ln{(q^2+4\mu^2)/\Lambda^2}}{\ln{(4\mu^2/\Lambda^2)}} 
        \right)^{-12/11}~,
\end{equation}
where $\Lambda$ is a scale and $\beta_0$ the leading coefficient of 
the QCD beta function.
The mass $\mu(q^2)$ is a \emph{dynamical} constituent gluon mass 
that vanishes as $q^2\to\infty$. The computation of a physical 
observable of known value then allows, in principle, to solve in
the coefficient $\mu$. In Ref.~\cite{Cornwall:1982zn}, a string 
inspired potential
$V(r)=2\mu(1-e^{-r/r_0})$, reminiscent of the one obtained in the
Schwinger model, is considered, where $\sigma_a=2\mu/r_0$.
Its effects are supplemented with a potential obtained
from the QCD $O(g^2)$ (now massive) one-gluon-exchange 
scattering amplitude. Fixing $m\simeq 0.5\,\GeV$, one obtains
$m(0^{++})\simeq1.2\,GeV $, $m(0^{-+})\simeq1.4\,\GeV$ 
and $m(2^{++})\simeq1.6\,\GeV$.
Note that the value of $\mu$ can be fixed in many different
ways, also owing to the fact that gluons are never observed. 
In Ref.~\cite{Bernard:1981pg}, for example, it is
defined as half the energy stored in a flux-tube between two static
sources trasforming in the adjoint representation of the gauge
group. One obtains, from Monte Carlo simulations of the lattice
regularized theory at finite lattice spacing and in the strong
coupling regime, $\mu\simeq 0.52\,\GeV$. 
The above is an admittedly 
very simplified picture, mainly because the system is treated 
non-relativistically. In a semi-relativistic treatment, the non
relativistic kinetic energy $p^2/2\mu$ is replaced with 
$\sqrt{p^2}$. A recent calculation, Ref.~\cite{Mathieu:2008bf} 
operates this improvement, which allows setting $\mu=0$, and 
also considers the contributions to the potential induced by 
instantons that should affect differently
states of different parity. As a result, $m(0^{++})=1.724\,\GeV$,
$m(0^{-+})=2.624\,\GeV$ and $m(2^{++})=2.588\,\GeV$ which, as we will
see, is in good agreement with the Lattice results.

A phenomenological and yet fully relativistic model is the MIT 
Bag Model, see Ref.~\cite{Chodos:1974je}, which was successful
in providing a qualitative understanding of many different properties 
of hadrons in terms of just a few parameters.
In this model, a hadron is a finite region 
of space of energy density $B$, and its internal
structure is described by quark and gluon fields. Confinement is introduced
by imposing vanishing boundary conditions on fields on the boundaries 
of the bag. 
In Ref.~\cite{Jaffe:1975fd}, the bag is approximated as a static sphere 
of radius $R$ and two families of eigenmodes of the free gluon field 
are found: the transverse electric($TE$) and transverse magnetic($TM$)
modes of energy $E=x_i/R$, where $i=TE$ or $TM$. Their quantum numbers
are easily determined as $x_{TE}=2.744$ and $J^{J+1,C}$ for $TE$ modes,
$x_{TM}=4.493$ and $J^{J,-}$ for $TM$ modes.
In Ref.~\cite{Johnson:1975zp}, the spetrum of glueballs was obtained by
populating the bag with $TE$ and $TM$ modes. 
The low-lying glueball 
states are found for $2$ or $3$ gluons, in agreeement with the qualitative
picture introduced at the beginning of this section. 
For $2$ gluons states 
we have $(TE)^2$ and $(TM)^2$ states, with $J^{PC}=0^{++},\, 2^{++},\ldots$.
and $(TE)(TM)$ states with $J^{PC}=0^{-+},\,2^{-+},\,\ldots$.
For $3$-gluon states we have $(TE)^3$ states with 
$J^{PC}=0^{+-},\, 1^{+-},\, 1^{--},\, 3^{+-},\ldots$.
Note the absence of any $1^{-+}$ state, 
which was argued, in Ref.~\cite{Rebbi:1975ns}, to describe 
the translation of the bag. 
As a consequence, itself and its contributions to other combinations 
should be discarded. This allows to exclude several states among
the $2$ gluon family that 
would be excluded, in the constituents models discussed previously, 
on the basis of Landau-Yang argument. 
The three lightest states are then found in the $0^{++}$, $2^{++}$
and $0^{-+}$ channels, with $m(0^{++})=m(2^{++})=0.96\,GeV$ and 
$m(0^{-+})=1.29\,GeV$. In Ref.~\cite{Chanowitz:1982qj}, the effect
of a (running) coupling is introduced between the modes. At 
leading order in the coupling
in a static cavity, the masses are found to be $m(0^{++})=0.67\,GeV$, 
$m(2^{++})=1.75\,GeV$ and $m(0^{-+})=1.44\,GeV$. For a non-static bag,
the eigenvalues of its hamiltonian must be related to the masses of the hadrons. 
In Ref.~\cite{Carlson:1982er} the effects of this center-of-mass motions
are taken into account and the bag constant $B$ is computed from a model
of the QCD vacuum. This leads to $m(0^{++})=1.58\,GeV$, 
$m(2^{++})=1.88\,GeV$ and $m(0^{-+})=0.81\,GeV$.

In Refs.~\cite{Isgur:1983wj,Isgur:1984bm}, a model of hadrons is defined from
the strong coupling limit Hamiltonian of lattice QCD. An analysis of the latter
reveals that in addition to the mesons and baryons of the ordinary 
quark model, its Hilbert space also contains glueballs, hybrids and other
exotic states. In the sector with no quarks, excitations are generated from
the vacuum by products of link operators on closed lattice paths. While
the states in strong coupling limit are not realized in continuum QCD, it is
argued that they form a complete basis of its Hilbert space. Hence, glueball
states are superposition of states generated by Wilson loops, and can be
described in the continuum by a non-relativistic model of a vibrating 
(circular) ring of glue. The low-lying spectrum of excitations 
yields $m(0^{++})=1.52\,GeV$, $m(0^{-+})=2.79\,GeV$, and 
$m(2^{++})=2.84\,GeV$. 

\begin{figure}
    \centering
    \scalebox{0.7}{\input{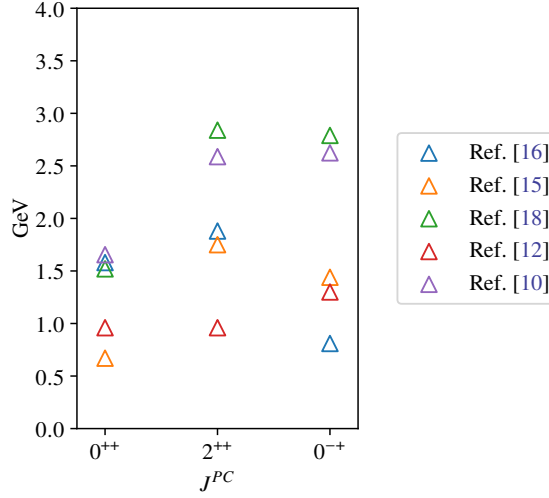}}
    \caption{The predictions on the spectrum of the $0^{++}$(scalar),
        $0^{-+}$(pseudo-scalar) and $2^{++}$(tensor) glueballs, 
        from the phenomenological models reviewed 
    in Section~\ref{sec:pheno}.\label{fig:pheno}}
\end{figure}

A summary of the above predictions for the spectrum of 
glueballs with quantum numbers $0^{++}$, $0^{-+}$ and 
$2^{++}$ channels is displayed in Figure~\ref{fig:pheno}.

\section{Analytical approaches} \label{sec:analytic}

In this section, two approaches based directly on the QCD 
Lagrangian are reviewed.
The first is based on Shifman-Vanshtein-Zakharov (SVZ) sum 
rules and the second is based on Bethe-Salpeter equations (BSE) 
for multi-gluon bound states.

SVZ sum rules, see Refs.~\cite{Shifman:1978bx,Shifman:1978by} 
and Ref.~\cite{Shifman:1998rb} 
for a pedagogical review, allow us to improve our understanding
of the non-perturbative regime of QCD. Measurable quantities 
like masses and decay constants of hadrons can be quantitatively related 
to the expectation values of local combinations of quark and gluon 
operators, 
known as \emph{condensates}. The condensates 
encode the long range properties of the QCD vacuum that 
are beyond the reach of perturbation theory. They cannot be calculated,
but after their values are fixed from phenomenology, predictions
can be formulated. The method of SVZ sum rules was very 
succesful in evaluating the 
spectrum and decay rates of ordinary mesons and baryons. 
The subject of the sum rules are the time ordered products at 
momentum $q$, 
\begin{equation}
    \Pi(q^2) = \imath \int\mathrm{d}^4 x~
    e^{\imath q\cdot x}
    \vevT{ J(x) J(0) }~,
\end{equation}
where the interpolating currents $J$ generate glueball states 
with the desired quantum numbers from the vacuum. For the 
scalar, pseudo-scalar and tensor glueballs, the currents are, 
\begin{align*}
	J_{0^{++}} &= \alpha_s \Tr G_{\mu\nu} G^{\mu\nu}\\
	J_{0^{-+}} &= \alpha_s \epsilon^{\mu\nu\rho\sigma}
    \Tr \tilde{G}_{\mu\nu} G_{\rho\sigma}\\
    J_{2^{++}}^{\mu\nu} &= -\Tr G^\mu_{\rho} G^{\nu\rho}
	+\frac{g_{\mu\nu}}{2} \Tr G_{\beta\alpha} G^{\beta\alpha}~,
\end{align*}
note the presence of $\alpha_s$ the strong coupling constant.
This can be calculated in perturbation theory at 
$Q^2=-q^2\gg\Lambda^2_\mathrm{QCD}$ and can be related, 
through the optical theorem, to the spectral
density $\rho(s) = \Im \Pi(q)/\pi$ for $s=q^2>0$, of states generated
by the current $J$. 
These two different regimes are related by the dispersion relation
\begin{equation}\label{eq:SVZSR}
    \Pi(q^2) = \frac{1}{\pi} 
    \int_{s_X}^\infty \mathrm{d} s \frac{\rho(s)}{s-(q^2+i0)} + 
    P(q^2)~,
\end{equation}
where $s_X$ is location of the first singularity of $\Pi(s)$ on the
real axis, and $P(q^2)$ is a polynomial that containts the subtractions
that are necessary when $\Pi(s)$ is divergent for $s\to\infty$.

The sum rules are used as follows. An ansatz is made for the spectral density, 
that contains a simple physical picture that captures our expectations
for the sector related to current $J$ and that contains the target 
observables quantities. The usual ansatz is
\begin{equation}
    \rho(s) = \frac{1}{\pi} f_X^2~\delta( s- m_X^2) + 
    \theta(s-S)\Im \Pi^\mathrm{QCD}(s)~,
\end{equation}
where $m_X$ is the mass of state $X$, 
$f_X=\langle 0|J(0)|X\rangle$ is its decay 
constant, and $S>m_X^2$ is the threshold of
energies over which the spectral density 
can be approximated by the perturbative one.
In this regime, $\Pi(q^2)$ can be calculated in terms
of quark and gluon fields at leading order, while the 
longer-range subleading contributions are 
obtained through the Operator Product Expansion (OPE). 
The condensates of appropriate local operators that appear in the OPE
may be classified according to their mass dimension $d$. For example,
for the scalar channel, the leading condensate, 
$\langle \Tr G_{\mu\nu} G^{\mu\nu}\rangle$, 
appears at $d=4$. The contribution of higher dimensionsional 
operators can be included, and becomes quantiatively relevant at smaller 
values of $Q^2$. 
In order to magnify the relative importance of the low-lying 
states, ideally in an energy range $Q^2\sim1\,GeV$, 
in the spectral density, and to suppress the effects of 
higher powers of $1/Q^2$ in
the OPE, a Borel transformation is usually performed on both sides
of the sum rule, Eq.~(\ref{eq:SVZSR}).  
Matching this computation with the ansarz for the spectral density
allows to relate $f_X$ and $m_X$ to phenomenology and to determine
their values, which can then be used to predict other quantities.
Clearly, the final estimates of 
$m_X$ and $f_X$ depend on which configurations are used to compute
$\Pi(Q^2)$ at large $Q^2$, for example whether instantons are included,
and on the value of the condensates.

In Refs.~\cite{Novikov:1979ux, Novikov:1979va}, 
the scalar and pseudoscalar currents were analyzed. The related glueballs
were put in correspondence with the $\eta'$ state at $1\,GeV$ 
and the $\sigma$-meson at $0.7\,GeV$. The contribution of instantons was
discussed and either not or only schematically taken into account.
Differently from, i.e. the $\rho$-meson, the contribution of instantons
at energy scales around $1\,GeV$ seems non-negligible, and
it is suggested that its neglect will impact
on the prediction of the glueball masses.
The matter is carefully analyzed in Ref.~\cite{Narison:1984hu,Narison:1996fm}, 
in which it is instead argued that the instanton contribution can be 
neglected. The masses of the scalar, pseudo-scalar and tensor
glueballs are predicted to be 
$m(0^{++})=1.5(2)\,GeV$, $m(0^{-+})=2.05(19)\,GeV$ and 
$m(2^{++})=2.0(1)\,GeV$.
In contrast, the effect of instantons is considered in 
in Ref.~\cite{Schafer:1994fd} and the direct instanton contribution is
evaluated in Ref.~\cite{Forkel:2000fd}. The calculation allows to predict 
the mass of the scalar glueball as $m(0^{++}) = 1.53(2)\,GeV$. A more recent
and systematic calculation of the contribution of the direct
instanton contribution may be found in in Ref.~\cite{Forkel:2003mk}.
The masses of the scalar, pseudoscalar, which are the most affected
are estimated as $m(0^{++})=1.25(2)\,GeV$, $m(0^{-+})=2.2(2)\,GeV$
In Ref.~\cite{Chen:2021bck,Chen:2022imp}, the authors
analyze very carefully the set of currents to correlate, and include 
condensates up to dimension 8, but not the contribution from instantons.
The masses of scalar, pseudoscalar and tensor glueballs 
are obtained as $m(0^{++}) = 1.78(17)\,GeV$, 
$m(0^{-+}) = 2.17(11)\,GeV$ and $m(2^{++}) = 1.86(17)\,GeV$.

\begin{figure}
    \centering
    \includegraphics[scale=0.4]{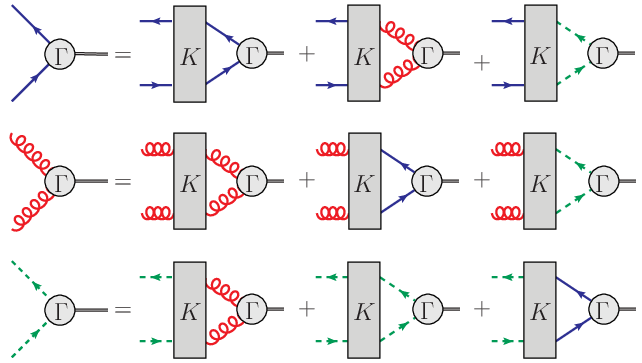}
  \caption{The coupled set of BSEs for two-body bound states
      of QCD. The blue, wiggly and dashed lines are the propagators
      for the quarks, gluons and ghots, respectively. The circles 
      are the Bethe-Salpeter amplitudes $\Gamma$ and the boxes
      the scattering kernels $K$. Taken 
        from Ref.~\cite{Huber:2022dsn}.\label{fig:BSE}}
\end{figure}

A different approach consists in using the 
Bethe-Salpeter formalism. In principle, this allows to obtain
information on bound states in a fully relativistic and 
non-perturbative manner. In practice, an infinite hierarchy of 
equations is involved, and approximations are needed to obtain
results.
In the Bethe-Salpeter equation (BSE), see Figure~\ref{fig:BSE}, 
one is interested in computing the amplitude $\Gamma$, given
an ansatz for the the two-body irreducible scattering kernel $K$
and the form of quark, gluon and ghost propagators.
In the absence of exact solutions for the latter, some kind of 
is needed, whose eventual effects are generally hard to control. 
Clearly, the choice of truncation and of the propagators
becomes the crucial aspect in determining the solidity of this method.

The first attempt at the calculation of the $J=0$ glueball in the BSE
framework can be found in Ref.~\cite{Meyers:2012ka}.
Available lattice results
on the behaviour of the gluon and ghost $2$-point function are used both to model
vertices through their Schwinger-Dyson equations and to ensure the correctness
of the resulting solution. These vertices are then used in truncated BSEs to 
predict the properties of glueballs. 
Assuming that the dressed version of the lowest order 
scattering kernel dominates the interaction and taking as input the 
mass of the scalar glueball computed in lattice simulations, 
the mass of the pseudoscalar glueball was obtained as $2.500(250)~GeV$.
In Ref.~\cite{Huber:2020ngt,Huber:2021yfy,Huber:2022dsn}, the BSEs were solved in 
in the Landau-gauge, in the pure Yang-Mills case. The truncation scheme
at $3$-loops was consistent between the DSE for $2$-points functions 
and vertices, obtained from the $3$PI effective action. There is 
no external parameter dependency apart from an overall scale. The latter
can be fixed by comparison with lattice results.
The scalar and pseudoscalar glueball masses were estimated as 
$m(0^{++})=1.850(130)\,GeV$,$m(0^{-+})=2.580(180)\,GeV$ and  
$m(2^{++})=5.610(180)\,GeV$.

The large-$N_c$ approach plays an important role in that it allows
to relate and combine results obtained at different values of $N_c$ with
the phenomenologically relevant case $N_c=3$. It is based on the 
observation that the calculation of amplitudes in Yang-Mills
theories drastically simplify when the number of colors $N_c$ is taken
to infinity keeping $g^2N_c$ fixed. In particular, if the large-$N_c$ theory 
is a confining theory, then it describes stable and non-interacting mesons
and glueballs, as can be easily understood from the scaling properties of 
Feynman diagrams with $N_c$.
At $N_c$ large but finite, it is possible to show that,
\begin{equation}
\frac{m(J^{PC})}{\sqrt{\sigma}} = m(N_c=\infty) + \frac{c_1}{N_c^2}~,
\end{equation}
where the coefficient $c_1$ is independent of $N_c$. This approach rests
on the possibility of computing the value of $m(N_c=\infty)$ and of $c_1$,
which can be achieved in several different ways. For a lattice oriented 
review, see Ref.~\cite{Lucini:2012gg}. Related approaches 
have been adopted in the context of the Ads/CFT correspondence. They
differ in the specific duality chosen, in the way the 
breaking of conformal symmetry is implemented, and in the 
identification of glueball operators. In the recent Ref.~\cite{Rinaldi:2021dxh},
the masses of the scalar and tensor glueballs are obtained in the
context of the graviton soft wall model, in which the glueball is associated
to a graviton propagating in $Ads_5$ space. The estimates of their masses
are $m(0^{++})=1.920\,\GeV$ and $m(2^{++})=2.371\,\GeV$. For further results,
see Section III.E of Ref.~\cite{Mathieu:2008me}.

\begin{figure}
    \centering
  \scalebox{0.7}{\input{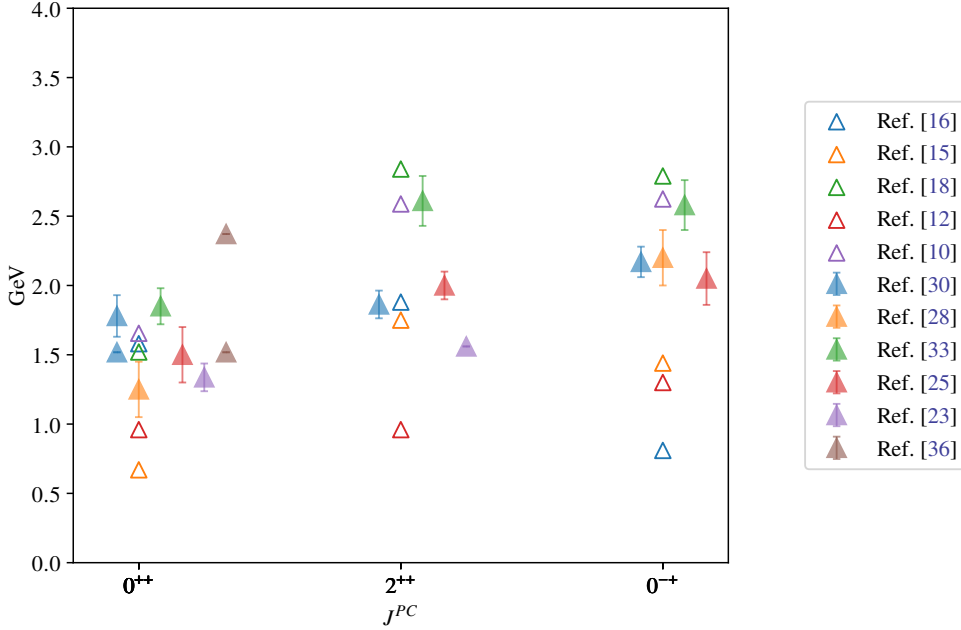}}
  \caption{The predictions on the spectrum of the $0^{++}$(scalar),
     $0^{-+}$(pseudo-scalar) and $2^{++}$(tensor) glueballs, 
     from the phenomenological models reviewed in Section
     ~\ref{sec:pheno} (empty triangles) and 
     from the analytical models reviewed 
     in Section~\ref{sec:analytic} (full triangles).\label{fig:analytic}}
\end{figure}

A summary of the above predictions for the spectrum of 
glueballs with quantum numbers $0^{++}$, $0^{-+}$ and 
$2^{++}$ channels is displayed in Figure~\ref{fig:analytic}.

\section{Lattice calculations} \label{sec:lat}

The numerical approach to lattice regularized quantum field 
theories is the only first principles approach to the exploration of 
the non-perturbative regime of QCD. As such, it is the instrument of
choice for the study of glueballs, in particular their
spectrum and the decay widths. In this section, estimates of 
the glueball spectral observables as theyr are usually obtained on the lattice 
are reviewed and the results present in the litterature are discussed.

The mass of a glueball state can be calculated from the large euclidean-time
behaviour of correlators of operators with appropriate quantum numbers. 
For zero-momentum projected operators, under very broad assumptions,
\begin{equation}\label{eq:correlator}
    C(t) = \langle \Omega | O(t) O^\dag(0) |\Omega\rangle 
    = \sum_n |c_n|^2 e^{-m_n t}~,
\end{equation}
where $n$ labels the eigenstate of the Hamiltonian, 
the quantities $|c_n|^2=|\langle n|O_i(0)|\Omega\rangle|^2$ 
are known as overlaps, and $m_n$ are the masses in the channel 
with the same quantum numbers as the operator $O$.
If there exists an isolated ground state of mass
$m_0$ then, at sufficiently large $t$, the sum in Eq.~\ref{eq:correlator}
will be dominated by $|c_0|^2 \exp{(-m_0 t)}$. In principle, the 
mass can be obtained as,
\begin{equation}
    m_0 = -\lim_{t\to\infty} \frac{1}{t}\log{ C(t) }~.
\end{equation}
In practice, the computation of $m_0$ with the above form for $C(t)$ at finite
values of $t$ will be affected by the contamination of higher energy states.
The average value above can be computed on the lattice as an ensemble 
average and is defined schematically as,
\begin{equation}
    C(t)= \frac{1}{Z} \int  \mathcal{D}[U] 
    \det M[U]\, O(t) O^\dag(0) \,e^{-S_\mathrm{YM}[U]}~,
\end{equation}
where $M[U]$ is the fermion matrix, $S_\mathrm{YM}[U]$ is the action for
the gluon field, and
\begin{equation}
    Z = \int  \mathcal{D}[U] \det M[U] e^{-S_\mathrm{YM}[U]}~.
\end{equation}
Many different choices are possible for both $S_\mathrm{YM}[U]$ and $M[U]$. 
For example, both isotropic and anisotropic discretizations can be
defined, and different actions characterized by different discretization
errors. 

At finite lattice spacing the states transform in irreducible representations
of the symmetries of the system. These are known as \emph{channels}.
The channels are labelled by $R^{PC}$, where $R$ are irreducible 
representations of the octahedral group $O_h$, $P$ is spatial 
parity and $C$ is charge conjugation. There are $10$ possible channels,
denoted by $A_1^\pm, A_2^\pm, E^\pm, T_1^\pm, T_2^\pm$. Their relationship
with the continuum channels $J^{PC}$, which they become part of in 
the continuum
limit, can be found in Table~\ref{tab:irreps}.
States are generated from the (invariant) vacuum $|\Omega\rangle$
by gauge-invariant combinations of link variables and quark fields.
Two families of such operators are known: traces of path-ordered products
along closed lattice paths $U_\mathcal{C}=\Tr~\prod_{l\in\mathcal{C}}U_l$,
and operators involving $q$ and $\bar{q}$ fields, $\bar{q} U_\mathcal{L} q$,
where $\mathcal{L}$ is a path connecting $\bar{q}$ and $q$.
The channel to which one operator belongs is dictated by the 
transformation properties of its support under elements 
of the octahedral group. As Charge conjugation simply amount to inverting
the ordering of the link operators along a path, the representations
with definite values of $C$ are simply obtained by considering 
the real and imaginary parts of ech $R^{P}$ representations. 

\begin{table}
    \centering
 \begin{tabular}{c|ccccc}
    $J$ & $A_1$ & $A_2$ & $E$ & $T_1$ & $T_2$ \\
    \hline
    $0$ &  $1$  &  $0$  & $0$ &  $0$  &  $0$  \\
    $1$ &  $0$  &  $0$  & $0$ &  $1$  &  $0$  \\
    $2$ &  $0$  &  $0$  & $1$ &  $0$  &  $1$  \\
    $3$ &  $0$  &  $1$  & $0$ &  $1$  &  $1$  \\
    $4$ &  $1$  &  $0$  & $1$ &  $1$  &  $1$  
  \end{tabular}
  \caption{In the top row, the representations of the octahedral 
      group. In the left column, the $J\leq4$ representations of the 
        Poincar\'e group. The elements $1$ of the matrix correspond
        to representations $R^{PC}$  that become part 
    of representation $J$ in the continuum limit.\label{tab:irreps}}
\end{table}

\begin{figure}
    \centering
   \includegraphics[scale=0.35]{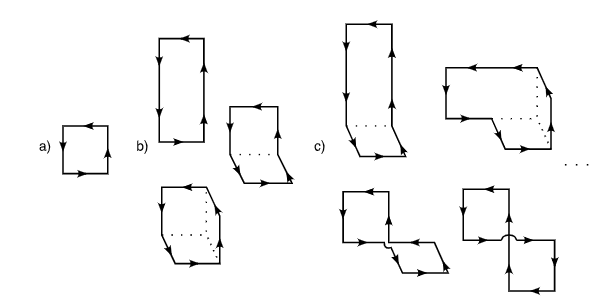}
   \caption{A sample of the lattice paths on which the glueball
       operators are defined. Taken from Ref.~\cite{Lucini:2010nv}.
\label{fig:op_variational}}
\end{figure}

It was soon realized that glueball correlators are 
affected by a particularly severe signal-to-noise ratio problem.
Two strategies have been proposed to overcome it. 
The first is based on the locality of both the Yang-Mills part of the
action and the objective operator.  It is known as multilevel, 
and allows to achieve an exponential reduction in the error of 
$C(t)$ at large $t$, see Refs.~\cite{Meyer:2002cd,DellaMorte:2010yp}. 
Unfortunately, it rests on locality, and its use is limited to quenched theories.
The second is known as \emph{variational method}~\cite{Wilson:1974sk,Ishikawa:1982tb}. A \emph{variational basis} of operators $\left\{O_i\right\}$ is 
defined in a given channel, and their correlation matrix 
is obtained,
\begin{equation}
    C_{ij}(t)=\langle \Omega | O_i(t) O_j(0) | \Omega\rangle
    = \sum_{n=1}^\infty c_{n,i} c_{n,j} e^{-m_n t}~,
\end{equation}
where $c_{n,i} = \langle n | O_i |\Omega\rangle$. By a diagonalization
of $C_{ij}(t)$ at large $t$, the ground state $m_0$ can in principle 
be obtained. In practice, because of the presence of 
statistical fluctuations, 
one instead solves the GEVP, $C(t) v = \lambda(t,\,t_0) C(t_0) v$, at
small $t$, where $v$ is a column vector, 
and $\lambda(t,\,t_0)=e^{-m_0(t-t_0)}$. This amounts to finding the
linear combination $\Phi(t)=\sum_i v_i O_i(t)$ of the operators that
maximize the overlap on the ground state in the channel. 
The mass can then be extracted from the large time behaviour of their
correlator. The great majority of the estimates of the spectrum have been
obtained using the variational method.
The efficacy of the method depends crucially on the choice of the
\emph{variational basis}. An sample of the closed loop operators
usually included is displayed in Figure~\ref{fig:op_variational}.
It has proved very
effective to add to the variational basis operators calculated on
blocked and smeared configurations~\cite{APE:1987ehd,Teper:1987wt,Lucini:2004my}.
This allows to better overlap with ground state configurations, especially
in the vicinity of the continuum limit, where they are
expected to be smooth on the $a$ scale. Moreover, as shown in 
Ref.~\cite{Lucini:2010nv}, the construction of the variational
basis can be automatized and the effect of scattering and di-torelon states
that propagate in the correlator can be identified. Estimates of the glueball 
masses can thus be 
obtained at several values of the inverse coupling $\beta$, 
and on several lattice geometries $N_s^2\times N_t$ and provided other
sources of systematicall error are addressed\footnote{For example, the loss 
of ergodicity caused by topological freezing, see below.}, an infinite volume 
continuum limit can then in principle be calculated. 

In quenched systems, where the fermionic degrees of freedom are infinitely 
massive and effectively static, the calculation of the glueball spectrum 
was one of the early successes of lattice QCD. It is nowadays
one of the best known results obtained on the lattice, and a solid prediction from
pure Yang-Mills theory. The spectrum can
be calculated from the Wilson action on isotropic lattices, 
see Ref.~\cite{Athenodorou:2020ani}, and from an improved action
on anisotropic lattices, see Refs.~\cite{Morningstar:1999rf,Chen:2005mg}. 
The high quality of these recent determinations of the spectrum rests
on the careful identification of the target states and on the quality of the
extrapolation to the continuum limit, and build upon decades of efforts. 

\begin{figure}
    \centering
   \includegraphics[scale=0.35]{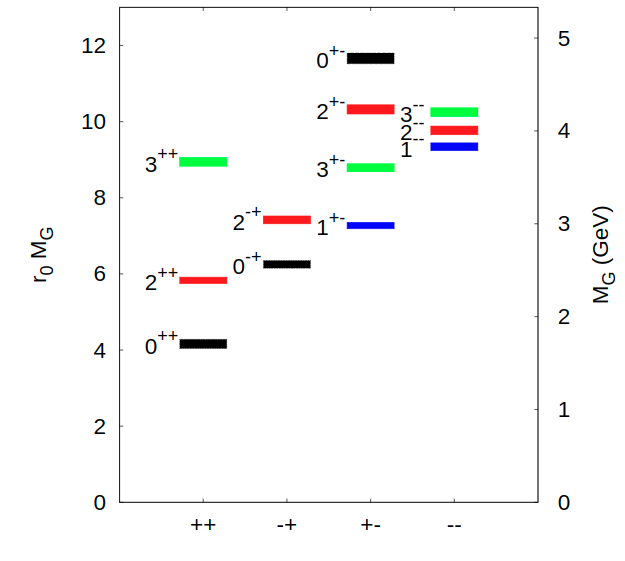}
   \scalebox{0.62}{\input{figures/spectrum_quenched_lattice.pgf}}
   \caption{(left) The quenched spectrum, taken from 
       Ref.~\cite{Chen:2005mg}. (right) A summary of the results
       obtained in the scalar, pseudo-scalar and tensor channels
       in quenched systems. The numerical values are taken 
       from Refs.~\cite{Morningstar:1999rf, Chen:2005mg, Athenodorou:2021qvs, Meyer:2004jc}.
\label{fig:quenched_spectrum}}
\end{figure}

The spectrum as obtained in Ref.~\cite{Chen:2005mg} is displayed in 
Fig.~\ref{fig:quenched_spectrum}.
The picture confirms the results obtained in the majority of 
models: the lightest channels are the scalar, followed by 
the tensor and the pseudo-scalar channels. The scalar glueball
in Ref.~\cite{Chen:2005mg} has a mass of $1.710(80)\,GeV$, 
the pseudoscalar a mass of
$2.560(120)\,GeV$ and the tensor $2.390(120)\,GeV$. In
Ref.~\cite{Athenodorou:2020ani} the scalar glueball has mass
$1.651(23)\,GeV$, the pseudoscalar $2.599(39)\,GeV$ and the
tensor $2.378(31)\,GeV$. These predictions are compatible with
each other within $1$-$\sigma$. Note that the choice of physical 
observables used to set the scale will have an effect on the 
final estimate in $GeV$ units. Estimates present in the litterature
are often expressed in units of the Sommer's scale $r_0$ or in units
of $\sqrt{\sigma}$. In recent investigations, results in units of
the Gradient Flow scale $t_0$ have started to appear.

Quenched systems are also an ideal testbed to investigate 
both the quantitative effects of other sources of systematical error
and to obtain the spectrum at larger values of $N_c$
or for different gauge groups.
For example, topological freezing affects simulations performed with periodic 
boundary conditions at small values of $a$. The resulting loss of 
ergodicity might affect mass estimations, especially in the pseudoscalar
channel. The problem has been analyzed at both $N_c=3$ and 
larger $N_c$, where the effect of topological freezing should be magnified, 
and found to be negligible, see Refs.~\cite{Chowdhury:2014kfa,
Chowdhury:2014mra,Amato:2015ipe,Bonanno:2022vot,Bonanno:2022yjr}.
Moreover, the spectrum was also evaluated at 
different values of the number of colors, see
Ref.~\cite{Lucini:2001ej, Athenodorou:2021qvs, Lucini:2010nv}, and 
extrapolated to the limit $N_c\to\infty$.
The lattice allows the exploration of the spectrum at large-$N_c$, 
see Ref.~\cite{Lucini:2010nv}
Finally, the glueball spectrum was obtained for gauge theories
based on other families of groups, see Ref.~\cite{Bennett:2020qtj} for
$Sp(N_c)$ gauge theories. The comparison of spectral data for
different families of gauge groups allows to analyze the degree of
universality among Yang-Mills theories. In particular, the Casimir 
scaling and the universality of the ratio between the tensor and scalar
glueball mass were analyzed in Refs.~\cite{Bennett:2020hqd,Hong:2017suj}.

A summary of the estimates of the spectrum in quenched lattice
QCD is displayed in the right-hand panel of Figure~\ref{fig:quenched_spectrum}.

The addition of dynamical fermions complicates the picture considerably.
The vacuum is altered in way that is difficult to predict.
Indeed, there is no reason to think that the unquenched and 
quenched spectra are similar: the presence of sea quarks of sufficiently
low mass makes glueballs unstable\footnote{Torelons, often used to measure
the scale, are affected by the same phenomenon.}, and the mixing with othere iso-singlet 
states makes it impossible to determine the 
very \emph{nature} of the state under scrutiny. In other words, a glueball mixed
with a $q\bar{q}$ component is indistinguishable from a meson with a glueball
component.
However, the possibility of varying the quark mass parameters smoothly in
Lattice calculations affords us a crucial advantage, in that it allows us to
tune the effects of the confounding phenomena above. 
In the regime of large quark masses, 
the system should be similar to its quenched limit, and the decays and mixing
effects should be inhibited. Well defined glueballs and quarkonia states are 
expected, and their mass shouyld be calculable with the methods above.
Reducing the sea quark mass smoothly reintroduces mixings and, 
for sufficiently small quark masses, decays. 

Early studies on the unquenched glueball spectrum have been performed
at $N_f=2$, with different fermion discretizations, in a regime of 
heavy quarks, see Refs.~\cite{Bitar:1991wr,Bali:2000vr,Hart:2001fp}. It is 
observed that, surprisingly, the statistical error in the determination
of the correlators is smaller than expected.
In Ref.~\cite{Bali:2000vr}, the masses of scalar and tensor glueballs at 
$N_f=2$ are obtained for Wilson fermions at $m_\pi\simeq0.490\,\GeV$ 
and are compared with quenched 
results for similar values of the lattice spacing. They are found to be 
in agreement within errors. 
In Ref.~\cite{Hart:2001fp}, a similar calculation is carried out for 
non-perturbatively improved clover Wilson fermions at 
$m_\pi\simeq 0.3-0.6\,\GeV$.
While the tensor glueball mass is found to be
in agreement with the quenched predictions, while in the scalar channel it
is found to be suppressed by $\sim 20\%$. As discussed
by the authors, as this discrepancy seem to be independent of the value
of the quark mass parameters, it might be explained as a lattice artifact
introduced by the $O(a)$ improvement.

\begin{figure}[h]
  \centering
  \includegraphics[scale=0.27]{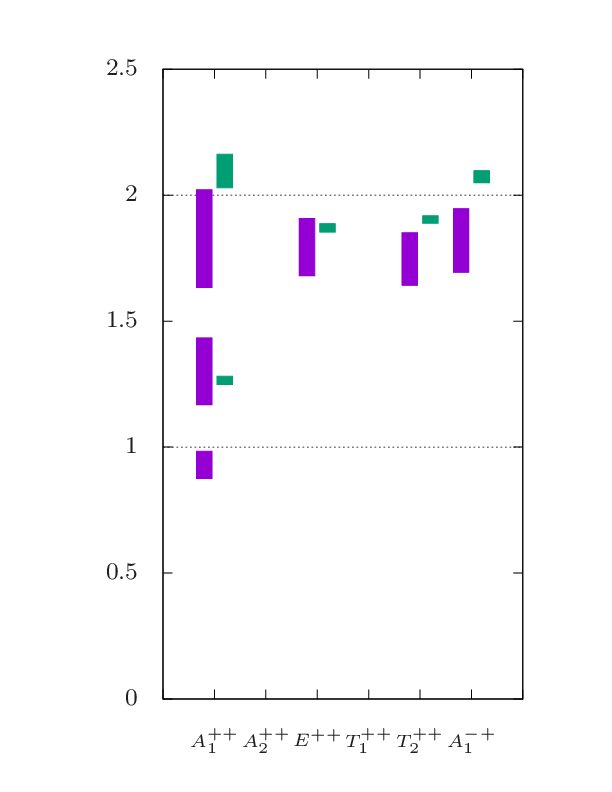}
    \includegraphics[scale=0.90]{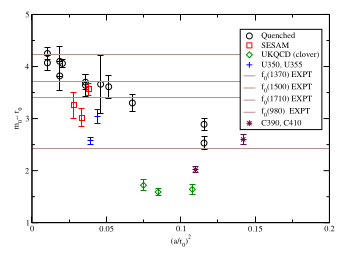}
  \caption{(left) In purple, the spectrum of glueballs 
      for $N_f=4$ clover improved 
	  twisted mass fermions, in green, the quenched spectrum from 
	  Refs.~\cite{Athenodorou:2021qvs,Athenodorou:2020ani} 
      $m_\mathrm{PS}\sim250~\mathrm{MeV}$. 
      Taken from Ref.~\cite{Athenodorou:2022nkb}.
      (right) Quenched and Unquenched spectrum of the scalar 
      glueball at finite $a$ and various values of $m_\pi$. The
      green diamonds are from Ref.~\cite{Hart:2001fp}, the
      blue and purple crosses from Ref.~\cite{Hart:2006ps},
      the red squares from Ref.~\cite{Bali:2000vr}. The quenched
      results from Ref.~\cite{Morningstar:1999rf} are
      represnted as black circles. Taken from Ref.~\cite{Hart:2006ps}.
\label{fig:unquenched_scalar}}
\end{figure}

The above results were obtained with only pure-glue operators in the variational
basis. In Ref.~\cite{McNeile:2000xx}, $q\bar{q}$ operators were included
in the variational basis and the mass of the 
flavor-singlet state was measured in a system
with $N_f=2$ flavors of clover improved Wilson fermions.
A suppression was observed in the mass of the 
flavor-singlet scalar state with respect to the results of 
Refs.~\cite{Bali:2000vr,Hart:2001fp}. 
Note that no difference was found between the mass
obtained from a purely gluonic variational basis and from a mixed one.
In order to interpret this suppression, the same calculation was
performed in Ref.~\cite{Hart:2006ps} in two different ways. 
First, using the same action but at a finer lattice, and also using
interpolating operators at non-zero momentum. Second, on ensembles 
generated by a \emph{gauge} improved Iwasaki action. In both cases, 
the suppression was still observed, see Figure~\ref{fig:unquenched_scalar}.
Two different interpretations thereof are put forward in 
Ref.~\cite{Hart:2006ps}. The suppression could be an artifact of the 
lattice discretization, caused by the so-called "scalar dip", or it 
could be a genuine effect of \emph{mixing}. 
The masses of glueballs in the scalar, pseudo-scalar and tensor
channels were later obtained for $2+1$ flavors of improved 
staggered fermions in Ref.~\cite{Richards:2010ck} and with a purely
gluonic variational basis. Only a weak 
dependence on the lattice spacing was found and the values of the
masses were found to be compatible with their value in the quenched 
continuum limit. The addition of scattering states to the variational 
basis did not alter this conclusion, see Ref.~\cite{Gregory:2012hu}.

More recently, these masses were computed using a purely gluonic
variational basis for an anisotropic lattice with $N_f=2$ 
clover-improved Wilson fermions in Ref.~\cite{Sun:2017ipk}. No
unquenching effects were detected on the masses of the 
pseudo-scalar and tensor channels. The scalar channel appears to have
a slightly suppressed mass with respect to its quenched counterpart.
However, no continuum limit is considered and more investigations
are needed before relating this effect to mixing.
The scalar glueball was the focus of Ref.~\cite{Brett:2019tzr}, where
for the first time, two-hadrons ($\pi\pi$, $K\bar{K}$ and $\eta\eta$), 
$q\bar{q}$ and purely-gluonic operators
were included in the variational basis. A single ensemble of $N_f=2+1$
clover improved Wilson fermions was analyzed at $m_\pi\sim 390\,MeV$,
using the stochastic LapH method to evaluate all-to-all quark propagation.
The aim of the authors was to understand whether, starting from the light
hadron spectrum obtained from only linear combinations of fermionic 
operators, additional states were observed to appear upon 
inclusion of glueball operators in the variational basis. Curiously,
no new state appears within the energy range considered. This is an
indication that further study is needed on the effects systematics
introduced by the choice of the variational basis.

\begin{figure}
    \centering
    \scalebox{0.7}{\input{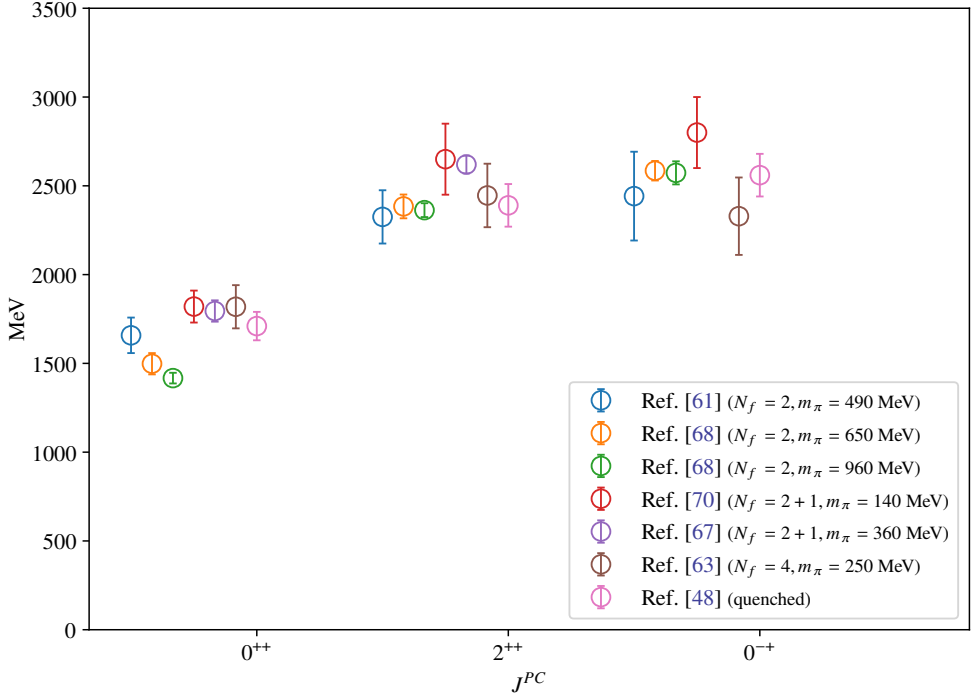}}
  \caption{A summary of estimates of the unquenched glueball
        spectrum. In light blue, the results from Ref.~\cite{Bali:2000vr},
    in light orange and green, the results from Ref.~\cite{Sun:2017ipk},
in red, the results in Ref.~\cite{Chen:2021dvn}, in purple the results
from Ref.~\cite{Gregory:2012hu}, in brown, the results from 
Ref.~\cite{Athenodorou:2022nkb}, in cyan the quenched results
from Ref.~\cite{Chen:2005mg}.\label{fig:unquenched_lattice}}
\end{figure}
At this conference, a calculation of the scalar glueball mass with $N_f=4$ 
clover improved twisted mass fermions was presented, 
see Ref.~\cite{Athenodorou:2022nkb}.
The low-quark mass regime was explored, with $m_\pi\sim250\,MeV$ and while in the
pseudo-scalar and tensor channel the masses were roughly found to agree with the 
corresponding quenched values, a new light state was observed in the 
scalar channel. Notably, the mass of the first and second 
excited states was found
to be similar to that the ground state and first excited quenched glueballs, 
respectively. The spectrum is displayed in in the left-hand panel of 
Figure~\ref{fig:unquenched_scalar}. It is
suggested that the new low-lying state is $\pi\pi$ or a $q\bar{q}$ state.
A similar calculation was performed for $N_f=2+1+1$. The fact that the
mass of the additional low-lying state was shown to depend strongly 
on $m_\pi$ suggests that it might contain a large quark content. 
The above results illustrate the need to improve our understanding
of the unquenched glueball spectrum, especially in the continuum
limit. However, the most pressing questions are on the effects of 
mixing.

A summary of the estimates of the spectrum in unquenched lattice
QCD at finite lattice spacing is displayed in Figure~\ref{fig:unquenched_lattice}.

The formalism to study the effects of mixing on the spectrum 
was described in detail in Ref.~\cite{McNeile:2000xx} where 
the mixing is found to be substantial. 
The same analysis was improved and obtained at smaller
lattice spacing in Ref.~\cite{Hart:2006ps}, with non-perturbatively
improved clover fermions at $a\sim 0.1\,fm$.
In Ref.~\cite{Richards:2010ck}, an approximated value of the 
non-diagonal element of the mixing matrix was obtained 
in a system of $N_f=2+1$ improved staggered fermions,
with $m_\pi\sim280\,MeV$ and $m_\pi\sim 360\,MeV$. While the
behaviour of the mixed correlation function was in agreement
with expectations, the data was still too noisy to draw any 
quantitative conclusion. Recently, the problem of computing
the disconnected contribution to the mixed correlators
has received some attention. In Ref.~\cite{Nino:2021klm,
Knechtli:2022bji}, the distillation method was used 
in a system with $N_f=2$ heavy quarks. 
In Ref.~\cite{Zhang:2021xvl}, $N_f=2$ flavors of
heavy quarks are considered on an anisotropic lattice. The mixing 
energy is computed from the explicit calculation of mixed 
glueball-quarkonia correlators. A mixing energy of $49(6)\,MeV$ 
and a mixing angle of $|\theta|=4.3(4)^\circ$ are obtained.
A similar study was performed in Ref.~\cite{Jiang:2022ffl}
using distillation to compute the values of the disconnected 
diagrams involved in the mixing dynamics.
A system of $N_f=2$ flavors of clover improved fermions on an
anisotropic lattice at $m_\pi\sim 350\,MeV$ was considered. 
The mixing energy was found to be $107(14)\,MeV$ and the 
mixing angle was estimated to be $|\theta|=2.47(46)^\circ$; 
this is small enough to argue that the mixing effects can be 
largely ignored in the pseudo-scalar channel
at $N_f=2$. 

A summary of the estimates of the spectrum in unquenched lattice
QCD at finite lattice spacing is displayed in Figure~\ref{fig:unquenched_lattice}.

\begin{figure}[h]
    \centering
  \includegraphics[scale=0.15]{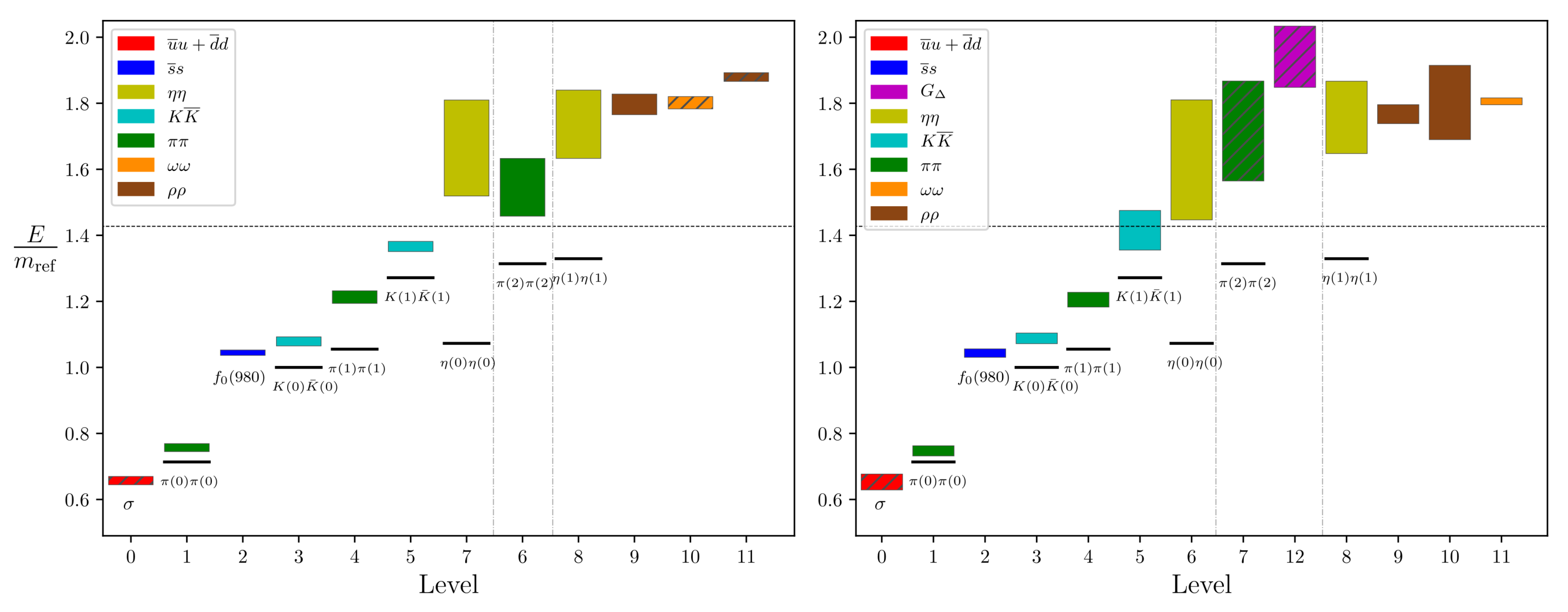}
  \caption{The spectrum of light hadrons without (left) and with (right)
      glueball operators in the variational basis, with $m_\mathrm{ref}=2m_\mathrm{K}$. Taken from Ref.~\cite{Brett:2019tzr}.}
\end{figure}

The prediction of the decay width of glueballs from QCD is crucial
to the comparison with experiments and is tightly related to the 
problem of determining the mixing between glueballs and other iso-singlet
states. A glueball with a very small decay width would 
already have been identified, while a very large resonance 
could remain out of reach forever. 

The calculation of decay widths
from lattice QCD is notoriously difficult and numerically very expensive. 
Despite the recent progresses in computing decay widths for mesons
using the L\"uscher-Lellouch
method, see Ref.~\cite{Lellouch:2000pv}, much remains to be 
done before this method can be fruitfully used for glueballs.
The main difficulty in the case of glueballs lies in the fact that,
since glueballs are iso-singlet states, studying their mixing
with $q\bar{q}$ state will necessarily involve the computation
of correlators corresponding to disconnected diagrams.
Decays glueballs to a pair of pseudo-scalar mesons were first studied 
in Refs.~\cite{Sexton:1994wg,Sexton:1995kd}. The $2$-point functions
of purely gluonic states with two-body meson operators at zero and 
non-zero value of the back-to-back momentum were studied in the quenched
approximation. The width of the decay to two pseudo-scalars 
was obtained as $108(29)\,MeV$ and the total decay width was estimate
to be smaller than $200\,MeV$, which would make the 
glueball well identifiable in experiments. 
Related to the decay of glueballs to other states is the decay of
the $J/\psi$ to glueballs. In Ref.~\cite{Gui:2012gx,Yang:2013xba,Gui:2019dtm},
the estimates of the decay widths to a photon plus a glueball 
are given for the scalar, pseudo-scalar and tensor channel in the 
quenched approximation and on anisotropic lattices, using the formalism
described in Ref.~\cite{Dudek:2006ej}.
A width of $0.35(8),\,keV$ is found, corresponding to a branching ratio
of $3.8(9)\times 10^{-3}$ for the scalar glueball. A width 
of $1.01(22)(10),keV$, corresponding to a branching 
ratio of $1.1(2)\times 10^{-2}$ for the
tensor glueball. A width of $0.0215(74),\,keV$, corresponding to a
branching ratio of $2.31(80)\times 10^{-4}$ for the pseudo-scalar glueball.
The study of the potential 
between glueballs and their scattering is a related and relevant problem, i.e.
for models of glueball dark matter. In this respect, 
see Refs.~\cite{Yamanaka:2021xqh,Yamanaka:2019yek}.

\section{Experimental results} \label{sec:exp}

The detection of glueball states is one of the long standing unsolved 
problems in hadron
spectroscopy. Several strategies have been developed in order to 
identify a signal that could
confirm their existence and a large number of studies have been 
conducted with that aim.
For recent review, see Ref.~\cite{Klempt:2007cp,Crede:2008vw,Chen:2022asf}. 

The typical experimental signature expected from a glueball is 
the appearance of supernumerary
states, that do not fit into the $q\bar{q}$ nonets of the minimal quark model. 
Their decay 
should exhibit branching fractions that are incompatible with those 
expected from $SU(3)$,
\begin{equation}
    \pi\pi:K\bar{K}:\eta\eta:\eta\eta' = 3:4:1:0~,
\end{equation}
and their 
detection should be best visible in Gluon-rich channels. The
latter are processes in which the not involving a glueball are 
suppressed, for example by the OZI-rule. 
The two confounding effects that have prevented the identification of a 
glueball signal so far are the mixing mentioned above on which, unfortunately,
little is known, and the possible presence of decay form factors.

\begin{figure}
    \centering
    \includegraphics[scale=0.35]{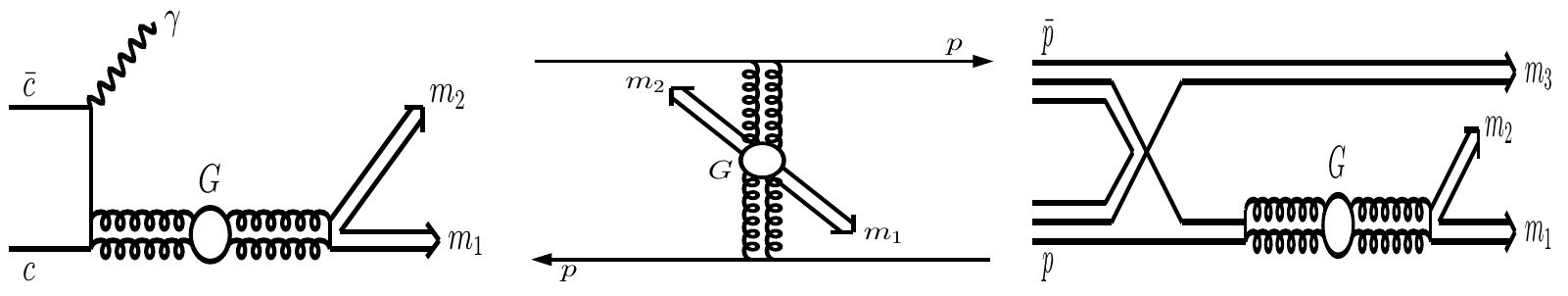}
    \caption{A sketch of three examples of Gluon rich processes. 
        In the left hand panel, the radiative
        decay $J/\psi\to\gamma m m$, where $m$ is a meson. In the
        centre panel, pomeron-pomeron collision, and in the right
        hand panel, hadron central production.
        Taken from Ref.~\cite{Klempt:2007cp}
        \label{fig:gluon_rich}}
\end{figure}

Three examples of gluon-rich channels that are currently under scrutiny 
in search of a glueball are displayed in Figure~\ref{fig:gluon_rich}. 
The $\bar{p}p$ annihilation is represented on the right-hand panel, 
a $q\bar{q}$ pair can annihilate and a glueball may be formed. This reaction
was studied at the Crystal Barrel 
experiment, see Ref.~\cite{CrystalBarrel:1992qav}, and will be the focus
of the future PANDA experiment at FAIR, see Refs.~\cite{Parganlija:2013xsa} for a
discussion involving the scalar glueball.
The diffractive scattering of hadrons is represented in the centre panel, 
and also known as double pomeron exchange. Since no valence 
quarks are exchanged, a glueball may form. This reaction was studied at the 
WA102 experiment. The radiative decay of the $J/\psi$ resonance is represented
on the left-hand panel. In this case, the OZI rule suppresses decay into 
light quarks, and the interesting process if the decay to a (detectable) photon 
and a pair of gluons. The pair of gluons might form a glueball, and its decay to
a pair of mesons might be identified. The BESIII experiment is currently
collecting data on this process.

Below, the focus is on the scalar sector, which is the most controversial. 
The number of states observed under $\sim 2.5\,GeV$ is still debated and so is
their identification. The states in this channel are listed in 
Table~\ref{tab:resonances}. Among those, $9$ are listed in the 2020 issue of the
PDG. The $f_0(500)$ and $f_0(980)$ have been mainly interpreted as molecular
states. The states $f_0(1370)$, $f_0(1500)$ where established by the Cristal Barrel
experiment through their decays to $\eta\eta$ and $\pi^0\pi^0$. Neither have
large coupling to $K\bar{K}$, and this would indicate that they cannot have a large
$s\bar{s}$ component. The results from the WA102 experiment are in agreement
with this view. The state $f_0(1710)$ was reported to decay predominantly to
$K\bar{K}$, indicating that it is predominantly $s\bar{s}$. 
Clearly, not all of these states can be scalar iso-scalars, 
and one of them must be supernumerary. 
The fact that experiments at LEP have indicated 
that the $f_0(1500)$ is practically absent from $\gamma\gamma\to K\bar{K}$ 
and $\gamma\gamma\to \pi^+\pi^-$ decays would suggest that it is predominantly
$s\bar{s}$, in contradiction with the above picture. The $f_0(1500)$ is thus
thought to be supernumerary. Yet the pattern of its decays to $\pi\pi$, $\eta\eta$,
$\eta'\eta'$ and $K\bar{K}$ indicate that it cannot be a pure glueball either. 
This points to the idea that the $f_0(1500)$ is a mixed state, partly glueball
partly $q\bar{q}$. 

Many studies revolve around the idea of mixing first explored in 
Ref.~\cite{Amsler:1995tu} for the scalar sector. Let $|G\rangle$, $|n\bar{n}\rangle$ and
$|s\bar{s}\rangle$, be the \emph{bare} states, where 
$|n\bar{n}\rangle = ( |u\bar{u}\rangle + |d\bar{d}\rangle)/\sqrt{2}$ and let
$M$ be the mass matrix,
\begin{equation}
M =
\begin{bmatrix}
   M_G & f & \sqrt{2} f \\
   f & M_S & 0 \\
   \sqrt{2}f & 0 & M_N
\end{bmatrix},\quad
      f=\langle s\bar{s} |V | G \rangle
       =\langle n\bar{n} |V | G \rangle/\sqrt{2}~,
\end{equation}
and $V$ is the potential generating the mixing. Then, the observed states, i.e.
the $f_0(1370)$, $f_0(1500$ and $f_0(1710)$ are the eigenvalues of $M$. 

Mixing also underlies the idea of a distributed glueball, put forward in 
Refs.~\cite{Sarantsev:2021ein,Klempt:2021wpg}. In this case, the various resonances in 
Table~\ref{tab:resonances} are ascribed either to a singlet or to a octet of $SU(3)$.
Their assignement can for example be decided from their Regge trajectories. 
While octet mesons should not appear in radiative decays of the $J/\psi$, 
they are abundantly produced. Moreover, singlets should be produced, but their yield
shows a peak at $\sim 1.865\,GeV$. This enhancement is interpreted as a scalar
glueball mixed into the wave-function of \emph{mainly}-octet and \emph{mainly}-singlet 
mesons.

\begin{figure}
    \centering
     \includegraphics[scale=0.35]{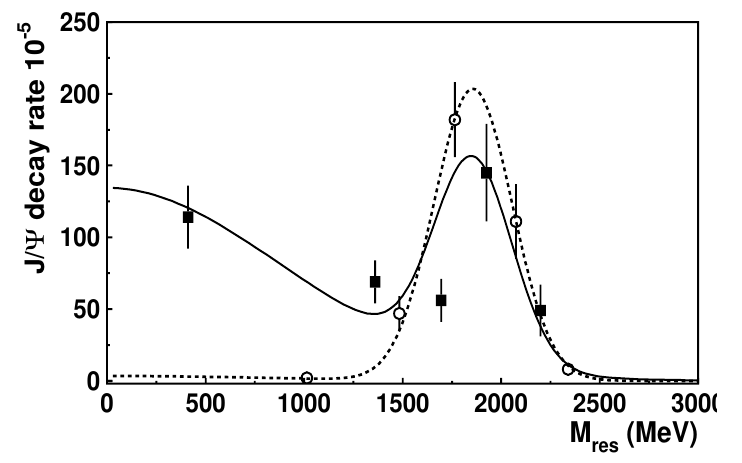}
     \includegraphics[scale=0.35]{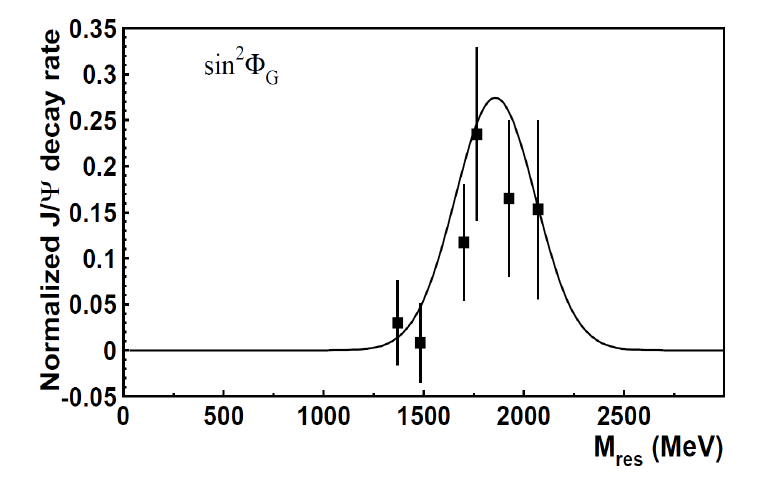}
     \caption{In the left-hand panel, the yield of mainly octet(circle) and 
         mainly-singlet(squares) mesons in the radiative $J/\psi$
         decays, plotted as a function of the mass $M_\mathrm{res}$
         of the resonance. In the right-hand panel, the glueball content
        of the resonances as a function of $M_\mathrm{res}$. The solid line
        is a Breit-Wigner distribution normalized to unit area, while the
        black squares are the values of $\sin^2\phi_n^s$, where
        $\phi_n^s$ is the mixing angle between non-glue states. 
        Note that the distribution of yields and the distribution of the glueball
        across the resonances is the same. Taken from Refs.~\cite{Sarantsev:2021ein}.}
\end{figure}
\begin{table}[h]
\centering
\begin{tabular}{lcccccc}
res. & $f_0(1370)$&$f_0(1500)$&$f_0(1710)$&$f_0(1770)$&$f_0(2020)$&$f_0(2100)$\\
G fraction & $5(4)\%$&$<5\%$&$12(6)\%$&$25(10)\%$&$16(9)\%$&$17(8)\%$
\end{tabular}
\caption{The glueball fraction found from a coupled-channel analysis of the decays
    of the $J/\psi$ into $\gamma\pi^0\pi^0$, $\gamma K_S K_S$, $\gamma\eta\eta$ and 
$\gamma\phi\omega$. Taken from Ref.~\cite{Sarantsev:2021ein}.\label{tab:glue_comp}}
\end{table}

In particular, a coupled channel analysis of the decays of $J/\psi$ into $\gamma\pi^0\pi^0$, 
$\gamma K_S K_S$, $\gamma\eta\eta$ and $\gamma\phi\omega$ indicates that a tenth state,
the $f_0(1770)$ is required to fit the invariant mass distributions of these decay products. 
The pole masses and width were obtained with the K-matrix approach, and the glueball 
contents of the resonances could be determined. They are reported in 
Table~\ref{tab:glue_comp}. Note that the fractions only amount to $\sim 78\%$
with a small fraction expected from $f_0(2020)$ and $f_0(2100)$. This indicates the presence
of a ``fractional glueball''
and to argue for a distributed scalar glueball with $m(0^{++}) = 1.865(50),\GeV$
and decay rate $\Gamma\sim 0.370(50)\,\GeV$.

\begin{table}
    \centering
   \begin{tabular}{l|cc}
     name & $m (\mathrm{MeV})$ & $\Gamma(\mathrm{MeV})$ \\
     \hline
     $f_0(500)\bullet$ & $400\to500$ & $480(30)$ \\
     $f_0(980)\bullet$ & $990(1)20$ & $71(10)$ \\
     $f_0(1370)\bullet$ & $1200\to1500$ & $200\to500$ \\
     $f_0(1500)\bullet$ & $1506(6)$ & $112(9)$ \\
     $f_0(1710)\bullet$ & $1704(12)$ & $123(18)$  \\
     $f_0(1770)$ & $1765(15)$ & $180(20)$ \\
     $f_0(2020)\bullet$ & $1992(16)$ & $442(60)$ \\
     $f_0(2100)\bullet$ & $2086^{20}_{-24}$ &  $284^{60}_{32}$\\
     $f_0(2200)\bullet$ & $2187(14)$ & $\sim 200$ \\
     $f_0(2330)\bullet$ & $\sim2330$ & $250(20)$
   \end{tabular}
   \caption{The masses and decay widths of the ten scalar iso-scalar 
   resonances considered in the text. The bullets indicate those
    that are confirmed in Ref.~\cite{10.1093/ptep/ptaa104}. The additional
    resonance $f_0(1770)$ was first proposed in Ref.~\cite{Bugg:2004xu}
    and is argued in Ref.~\cite{Sarantsev:2021ein} to be the tenth
    isoscalar resonance.\label{tab:resonances}}
\end{table}

\section{Conclusion}

Glueballs are among the predictions of QCD that still await undisputed confirmations. Many
different phenomenological models, more or less QCD inspired predict masses that are similar,
and in agreement with Lattice QCD results, obtained in the quenched approximation
\begin{equation}
    m(0^{++})= 1.6(1)\,GeV,\quad
    m(2^{++})= 2.4(1)\,GeV,\quad
    m(2^{++})= 2.5(1)\,GeV,\quad
\end{equation}
and $\Gamma\simeq 0.1\,GeV$ for the decay width. 

Unquenched results, that are numerically more expensive, 
are also affected by obscuring effects, mixing and decays, and a first principles 
study that includes a continuum 
extrapolation at the physical pion mass is still lacking. Decay widths have been calculated
but are affected by large systematical errors.

The experimental situation is debated. There is agreement on the present of a supernumerary
state in the scalar iso-scalar channel, but several scenarios, based on the idea of mixing,
are equally plausible for the identification of the scalar glueball. Further information, 
coming from the currently running BESIII experiment might prove useful.

\acknowledgments

The author would like to thank  B.~Lucini, A.~Athenodorou, V.~Drach, C.~McNeile, C.~Bonanno,
M.~Peardon, A.~Patella, G.~Bali, and F.~Knechtli for useful discussion.


    \bibliography{glueballs}

\begin{thebibliography}{10}

\bibitem{Fritzsch:1972jv}
Harald Fritzsch and Murray Gell-Mann.
\newblock {Current algebra: Quarks and what else?}
\newblock {\em eConf}, C720906V2:135--165, 1972.

\bibitem{Jaffe:1985qp}
R.~L. Jaffe, K.~Johnson, and Z.~Ryzak.
\newblock {Qualitative Features of the Glueball Spectrum}.
\newblock {\em Annals Phys.}, 168:344, 1986.

\bibitem{Landau:1948kw}
L.~D. Landau.
\newblock {On the angular momentum of a system of two photons}.
\newblock {\em Dokl. Akad. Nauk SSSR}, 60(2):207--209, 1948.

\bibitem{Yang:1950rg}
Chen-Ning Yang.
\newblock {Selection Rules for the Dematerialization of a Particle Into Two Photons}.
\newblock {\em Phys. Rev.}, 77:242--245, 1950.

\bibitem{Barnes:1981ac}
Ted Barnes.
\newblock {A Transverse Gluonium Potential Model With Breit-fermi Hyperfine Effects}.
\newblock {\em Z. Phys. C}, 10:275, 1981.

\bibitem{Cornwall:1981zr}
John~M. Cornwall.
\newblock {Dynamical Mass Generation in Continuum QCD}.
\newblock {\em Phys. Rev. D}, 26:1453, 1982.

\bibitem{Aguilar:2006gr}
Arlene~C. Aguilar and Joannis Papavassiliou.
\newblock {Gluon mass generation in the PT-BFM scheme}.
\newblock {\em JHEP}, 12:012, 2006.

\bibitem{Cornwall:1982zn}
J.~M. Cornwall and A.~Soni.
\newblock {Glueballs as Bound States of Massive Gluons}.
\newblock {\em Phys. Lett. B}, 120:431, 1983.

\bibitem{Bernard:1981pg}
Claude~W. Bernard.
\newblock {Monte Carlo Evaluation of the Effective Gluon Mass}.
\newblock {\em Phys. Lett. B}, 108:431--434, 1982.

\bibitem{Mathieu:2008bf}
Vincent Mathieu, Fabien Buisseret, and Claude Semay.
\newblock {Gluons in glueballs: Spin or helicity?}
\newblock {\em Phys. Rev. D}, 77:114022, 2008.

\bibitem{Chodos:1974je}
A.~Chodos, R.~L. Jaffe, K.~Johnson, Charles~B. Thorn, and V.~F. Weisskopf.
\newblock {A New Extended Model of Hadrons}.
\newblock {\em Phys. Rev. D}, 9:3471--3495, 1974.

\bibitem{Jaffe:1975fd}
R.~L. Jaffe and K.~Johnson.
\newblock {Unconventional States of Confined Quarks and Gluons}.
\newblock {\em Phys. Lett. B}, 60:201--204, 1976.

\bibitem{Johnson:1975zp}
K.~Johnson.
\newblock {The M.I.T. Bag Model}.
\newblock {\em Acta Phys. Polon. B}, 6:865, 1975.

\bibitem{Rebbi:1975ns}
C.~Rebbi.
\newblock {Nonspherical Deformations of Hadronic Bags}.
\newblock {\em Phys. Rev. D}, 12:2407, 1975.

\bibitem{Chanowitz:1982qj}
Michael~S. Chanowitz and Stephen~R. Sharpe.
\newblock {Hybrids: Mixed States of Quarks and Gluons}.
\newblock {\em Nucl. Phys. B}, 222:211--244, 1983.
\newblock [Erratum: Nucl.Phys.B 228, 588--588 (1983)].

\bibitem{Carlson:1982er}
C.~E. Carlson, T.~H. Hansson, and C.~Peterson.
\newblock {Meson, Baryon and Glueball Masses in the MIT Bag Model}.
\newblock {\em Phys. Rev. D}, 27:1556--1564, 1983.

\bibitem{Isgur:1983wj}
Nathan Isgur and Jack~E. Paton.
\newblock {A Flux Tube Model for Hadrons}.
\newblock {\em Phys. Lett. B}, 124:247--251, 1983.

\bibitem{Isgur:1984bm}
Nathan Isgur and Jack~E. Paton.
\newblock {A Flux Tube Model for Hadrons in QCD}.
\newblock {\em Phys. Rev. D}, 31:2910, 1985.

\bibitem{Shifman:1978bx}
Mikhail~A. Shifman, A.~I. Vainshtein, and Valentin~I. Zakharov.
\newblock {QCD and Resonance Physics. Theoretical Foundations}.
\newblock {\em Nucl. Phys. B}, 147:385--447, 1979.

\bibitem{Shifman:1978by}
Mikhail~A. Shifman, A.~I. Vainshtein, and Valentin~I. Zakharov.
\newblock {QCD and Resonance Physics: Applications}.
\newblock {\em Nucl. Phys. B}, 147:448--518, 1979.

\bibitem{Shifman:1998rb}
Mikhail~A. Shifman.
\newblock {Snapshots of hadrons or the story of how the vacuum medium determines the properties of the classical mesons which are produced, live and die in the QCD vacuum}.
\newblock {\em Prog. Theor. Phys. Suppl.}, 131:1--71, 1998.

\bibitem{Novikov:1979ux}
V.~A. Novikov, Mikhail~A. Shifman, A.~I. Vainshtein, and Valentin~I. Zakharov.
\newblock {eta-prime Meson as Pseudoscalar Gluonium}.
\newblock {\em Phys. Lett. B}, 86:347, 1979.

\bibitem{Novikov:1979va}
V.~A. Novikov, Mikhail~A. Shifman, A.~I. Vainshtein, and Valentin~I. Zakharov.
\newblock {In a Search for Scalar Gluonium}.
\newblock {\em Nucl. Phys. B}, 165:67--79, 1980.

\bibitem{Narison:1984hu}
Stephan Narison.
\newblock {Spectral Function Sum Rules for Gluonic Currents}.
\newblock {\em Z. Phys. C}, 26:209, 1984.

\bibitem{Narison:1996fm}
Stephan Narison.
\newblock {Masses, decays and mixings of gluonia in QCD}.
\newblock {\em Nucl. Phys. B}, 509:312--356, 1998.

\bibitem{Schafer:1994fd}
Thomas Sch\"afer and Edward~V. Shuryak.
\newblock {Glueballs and instantons}.
\newblock {\em Phys. Rev. Lett.}, 75:1707--1710, 1995.

\bibitem{Forkel:2000fd}
Hilmar Forkel.
\newblock {Scalar gluonium and instantons}.
\newblock {\em Phys. Rev. D}, 64:034015, 2001.

\bibitem{Forkel:2003mk}
Hilmar Forkel.
\newblock {Direct instantons, topological charge screening and QCD glueball sum rules}.
\newblock {\em Phys. Rev. D}, 71:054008, 2005.

\bibitem{Chen:2021bck}
Hua-Xing Chen, Wei Chen, and Shi-Lin Zhu.
\newblock {Two- and three-gluon glueballs of C=+}.
\newblock {\em Phys. Rev. D}, 104(9):094050, 2021.

\bibitem{Chen:2022imp}
Hua-Xing Chen, Wei Chen, and Shi-Lin Zhu.
\newblock {Two- and three-gluon glueballs within QCD sum rules}.
\newblock {\em Nucl. Part. Phys. Proc.}, 318-323:122--126, 2022.

\bibitem{Huber:2022dsn}
Markus~Q. Huber, Christian~S. Fischer, and Helios Sanchis-Alepuz.
\newblock {Glueballs from bound state equations}.
\newblock {\em EPJ Web Conf.}, 274:03016, 2022.

\bibitem{Meyers:2012ka}
Joseph Meyers and Eric~S. Swanson.
\newblock {Spin Zero Glueballs in the Bethe-Salpeter Formalism}.
\newblock {\em Phys. Rev. D}, 87(3):036009, 2013.

\bibitem{Huber:2020ngt}
Markus~Q. Huber, Christian~S. Fischer, and H\`elios Sanchis-Alepuz.
\newblock {Spectrum of scalar and pseudoscalar glueballs from functional methods}.
\newblock {\em Eur. Phys. J. C}, 80(11):1077, 2020.

\bibitem{Huber:2021yfy}
Markus~Q. Huber, Christian~S. Fischer, and Helios Sanchis-Alepuz.
\newblock {Higher spin glueballs from functional methods}.
\newblock {\em Eur. Phys. J. C}, 81(12):1083, 2021.
\newblock [Erratum: Eur.Phys.J.C 82, 38 (2022)].

\bibitem{Lucini:2012gg}
Biagio Lucini and Marco Panero.
\newblock {SU(N) gauge theories at large N}.
\newblock {\em Phys. Rept.}, 526:93--163, 2013.

\bibitem{Rinaldi:2021dxh}
Matteo Rinaldi and Vicente Vento.
\newblock {Meson and glueball spectroscopy within the graviton soft wall model}.
\newblock {\em Phys. Rev. D}, 104(3):034016, 2021.

\bibitem{Mathieu:2008me}
Vincent Mathieu, Nikolai Kochelev, and Vicente Vento.
\newblock {The Physics of Glueballs}.
\newblock {\em Int. J. Mod. Phys. E}, 18:1--49, 2009.

\bibitem{Meyer:2002cd}
Harvey~B. Meyer.
\newblock {Locality and statistical error reduction on correlation functions}.
\newblock {\em JHEP}, 01:048, 2003.

\bibitem{DellaMorte:2010yp}
Michele Della~Morte and Leonardo Giusti.
\newblock {A novel approach for computing glueball masses and matrix elements in Yang-Mills theories on the lattice}.
\newblock {\em JHEP}, 05:056, 2011.

\bibitem{Wilson:1974sk}
Kenneth~G. Wilson.
\newblock {Confinement of Quarks}.
\newblock {\em Phys. Rev. D}, 10:2445--2459, 1974.

\bibitem{Ishikawa:1982tb}
K.~Ishikawa, M.~Teper, and G.~Schierholz.
\newblock {The Glueball Mass Spectrum in {QCD}: First Results of a Lattice Monte Carlo Calculation}.
\newblock {\em Phys. Lett. B}, 110:399--405, 1982.

\bibitem{APE:1987ehd}
M.~Albanese et~al.
\newblock {Glueball Masses and String Tension in Lattice QCD}.
\newblock {\em Phys. Lett. B}, 192:163--169, 1987.

\bibitem{Teper:1987wt}
M.~Teper.
\newblock {An Improved Method for Lattice Glueball Calculations}.
\newblock {\em Phys. Lett. B}, 183:345, 1987.

\bibitem{Lucini:2004my}
Biagio Lucini, Michael Teper, and Urs Wenger.
\newblock {Glueballs and k-strings in SU(N) gauge theories: Calculations with improved operators}.
\newblock {\em JHEP}, 06:012, 2004.

\bibitem{Lucini:2010nv}
Biagio Lucini, Antonio Rago, and Enrico Rinaldi.
\newblock {Glueball masses in the large N limit}.
\newblock {\em JHEP}, 08:119, 2010.

\bibitem{Athenodorou:2020ani}
Andreas Athenodorou and Michael Teper.
\newblock {The glueball spectrum of SU(3) gauge theory in 3 + 1 dimensions}.
\newblock {\em JHEP}, 11:172, 2020.

\bibitem{Morningstar:1999rf}
Colin~J. Morningstar and Mike~J. Peardon.
\newblock {The Glueball spectrum from an anisotropic lattice study}.
\newblock {\em Phys. Rev. D}, 60:034509, 1999.

\bibitem{Chen:2005mg}
Y.~Chen et~al.
\newblock {Glueball spectrum and matrix elements on anisotropic lattices}.
\newblock {\em Phys. Rev. D}, 73:014516, 2006.

\bibitem{Athenodorou:2021qvs}
Andreas Athenodorou and Michael Teper.
\newblock {SU(N) gauge theories in 3+1 dimensions: glueball spectrum, string tensions and topology}.
\newblock {\em JHEP}, 12:082, 2021.

\bibitem{Meyer:2004jc}
Harvey~B. Meyer and Michael~J. Teper.
\newblock {Glueball Regge trajectories and the pomeron: A Lattice study}.
\newblock {\em Phys. Lett. B}, 605:344--354, 2005.

\bibitem{Chowdhury:2014kfa}
Abhishek Chowdhury, A.~Harindranath, and Jyotirmoy Maiti.
\newblock {Open Boundary Condition, Wilson Flow and the Scalar Glueball Mass}.
\newblock {\em JHEP}, 06:067, 2014.

\bibitem{Chowdhury:2014mra}
Abhishek Chowdhury, A.~Harindranath, and Jyotirmoy Maiti.
\newblock {Correlation and localization properties of topological charge density and the pseudoscalar glueball mass in SU(3) lattice Yang-Mills theory}.
\newblock {\em Phys. Rev. D}, 91(7):074507, 2015.

\bibitem{Amato:2015ipe}
Alessandro Amato, Gunnar Bali, and Biagio Lucini.
\newblock {Topology and glueballs in $SU(7)$ Yang-Mills with open boundary conditions}.
\newblock {\em PoS}, LATTICE2015:292, 2016.

\bibitem{Bonanno:2022vot}
Claudio Bonanno, Massimo D'Elia, Biagio Lucini, and Davide Vadacchino.
\newblock {Towards glueball masses of large-$N$ $\mathrm{SU}(N)$ Yang-Mills theories without topological freezing via parallel tempering on boundary conditions}.
\newblock {\em PoS}, LATTICE2022:392, 2023.

\bibitem{Bonanno:2022yjr}
Claudio Bonanno, Massimo D'Elia, Biagio Lucini, and Davide Vadacchino.
\newblock {Towards glueball masses of large-$N$ $\mathrm{SU}(N)$ pure-gauge theories without topological freezing}.
\newblock {\em Phys. Lett. B}, 833:137281, 2022.

\bibitem{Lucini:2001ej}
B.~Lucini and M.~Teper.
\newblock {SU(N) gauge theories in four-dimensions: Exploring the approach to N = infinity}.
\newblock {\em JHEP}, 06:050, 2001.

\bibitem{Bennett:2020qtj}
Ed~Bennett, Jack Holligan, Deog~Ki Hong, Jong-Wan Lee, C.~J.~David Lin, Biagio Lucini, Maurizio Piai, and Davide Vadacchino.
\newblock {Glueballs and strings in $Sp(2N)$ Yang-Mills theories}.
\newblock {\em Phys. Rev. D}, 103(5):054509, 2021.

\bibitem{Bennett:2020hqd}
Ed~Bennett, Jack Holligan, Deog~Ki Hong, Jong-Wan Lee, C.~J.~David Lin, Biagio Lucini, Maurizio Piai, and Davide Vadacchino.
\newblock {Color dependence of tensor and scalar glueball masses in Yang-Mills theories}.
\newblock {\em Phys. Rev. D}, 102(1):011501, 2020.

\bibitem{Hong:2017suj}
Deog~Ki Hong, Jong-Wan Lee, Biagio Lucini, Maurizio Piai, and Davide Vadacchino.
\newblock {Casimir scaling and Yang\textendash{}Mills glueballs}.
\newblock {\em Phys. Lett. B}, 775:89--93, 2017.

\bibitem{Bitar:1991wr}
Khalil~M. Bitar et~al.
\newblock {On glueballs and topology in lattice QCD with two light flavors}.
\newblock {\em Phys. Rev. D}, 44:2090--2109, 1991.

\bibitem{Bali:2000vr}
Gunnar~S. Bali, Bram Bolder, Norbert Eicker, Thomas Lippert, Boris Orth, Peer Ueberholz, Klaus Schilling, and Thorsten Struckmann.
\newblock {Static potentials and glueball masses from QCD simulations with Wilson sea quarks}.
\newblock {\em Phys. Rev. D}, 62:054503, 2000.

\bibitem{Hart:2001fp}
A.~Hart and M.~Teper.
\newblock {On the glueball spectrum in O(a) improved lattice QCD}.
\newblock {\em Phys. Rev. D}, 65:034502, 2002.

\bibitem{Athenodorou:2022nkb}
Andreas Athenodorou, Jacob Finkenrath, Adam Lantos, and Michael Teper.
\newblock {The glueball spectrum with $N_f=4$ light fermions}.
\newblock {\em PoS}, LATTICE2022:057, 2023.

\bibitem{Hart:2006ps}
A.~Hart, C.~McNeile, Christopher Michael, and J.~Pickavance.
\newblock {A Lattice study of the masses of singlet 0++ mesons}.
\newblock {\em Phys. Rev. D}, 74:114504, 2006.

\bibitem{McNeile:2000xx}
Craig McNeile and Christopher Michael.
\newblock {Mixing of scalar glueballs and flavor singlet scalar mesons}.
\newblock {\em Phys. Rev. D}, 63:114503, 2001.

\bibitem{Richards:2010ck}
Christopher~M. Richards, Alan~C. Irving, Eric~B. Gregory, and Craig McNeile.
\newblock {Glueball mass measurements from improved staggered fermion simulations}.
\newblock {\em Phys. Rev. D}, 82:034501, 2010.

\bibitem{Gregory:2012hu}
E.~Gregory, A.~Irving, B.~Lucini, C.~McNeile, A.~Rago, C.~Richards, and E.~Rinaldi.
\newblock {Towards the glueball spectrum from unquenched lattice QCD}.
\newblock {\em JHEP}, 10:170, 2012.

\bibitem{Sun:2017ipk}
Wei Sun, Long-Cheng Gui, Ying Chen, Ming Gong, Chuan Liu, Yu-Bin Liu, Zhaofeng Liu, Jian-Ping Ma, and Jian-Bo Zhang.
\newblock {Glueball spectrum from $N_f=2$ lattice QCD study on anisotropic lattices}.
\newblock {\em Chin. Phys. C}, 42(9):093103, 2018.

\bibitem{Brett:2019tzr}
Ruair\'\i{} Brett, John Bulava, Daniel Darvish, Jacob Fallica, Andrew Hanlon, Ben H\"orz, and Colin Morningstar.
\newblock {Spectroscopy From The Lattice: The Scalar Glueball}.
\newblock {\em AIP Conf. Proc.}, 2249(1):030032, 2020.

\bibitem{Chen:2021dvn}
Feiyu Chen, Xiangyu Jiang, Ying Chen, Keh-Fei Liu, Wei Sun, and Yi-Bo Yang.
\newblock {Glueballs at Physical Pion Mass}.
\newblock 11 2021.

\bibitem{Nino:2021klm}
Juan Andr\'es~Urrea Ni\~no, Francesco Knechtli, Tomasz Korzec, and Mike Peardon.
\newblock {Optimizing distillation for charmonium and glueballs}.
\newblock {\em PoS}, LATTICE2021:314, 2022.

\bibitem{Knechtli:2022bji}
Francesco Knechtli, Tomasz Korzec, Michael Peardon, and Juan~Andr\'es Urrea-Ni\~no.
\newblock {Optimizing creation operators for charmonium spectroscopy on the lattice}.
\newblock {\em Phys. Rev. D}, 106(3):034501, 2022.

\bibitem{Zhang:2021xvl}
Renqiang Zhang, Wei Sun, Ying Chen, Ming Gong, Long-Cheng Gui, and Zhaofeng Liu.
\newblock {The glueball content of \ensuremath{\eta}c}.
\newblock {\em Phys. Lett. B}, 827:136960, 2022.

\bibitem{Jiang:2022ffl}
Xiangyu Jiang, Wei Sun, Feiyu Chen, Ying Chen, Ming Gong, Zhaofeng Liu, and Renqiang Zhang.
\newblock {$\eta$-glueball mixing from $N_f=2$ lattice QCD}.
\newblock 5 2022.

\bibitem{Lellouch:2000pv}
Laurent Lellouch and Martin Luscher.
\newblock {Weak transition matrix elements from finite volume correlation functions}.
\newblock {\em Commun. Math. Phys.}, 219:31--44, 2001.

\bibitem{Sexton:1994wg}
J.~Sexton, A.~Vaccarino, and D.~Weingarten.
\newblock {Scalar glueball decay}.
\newblock {\em Nucl. Phys. B Proc. Suppl.}, 42:279--281, 1995.

\bibitem{Sexton:1995kd}
J.~Sexton, A.~Vaccarino, and D.~Weingarten.
\newblock {Numerical evidence for the observation of a scalar glueball}.
\newblock {\em Phys. Rev. Lett.}, 75:4563--4566, 1995.

\bibitem{Gui:2012gx}
Long-Cheng Gui, Ying Chen, Gang Li, Chuan Liu, Yu-Bin Liu, Jian-Ping Ma, Yi-Bo Yang, and Jian-Bo Zhang.
\newblock {Scalar Glueball in Radiative $J/\psi$ Decay on the Lattice}.
\newblock {\em Phys. Rev. Lett.}, 110(2):021601, 2013.

\bibitem{Yang:2013xba}
Yi-Bo Yang, Long-Cheng Gui, Ying Chen, Chuan Liu, Yu-Bin Liu, Jian-Ping Ma, and Jian-Bo Zhang.
\newblock {Lattice Study of Radiative J/\ensuremath{\psi} Decay to a Tensor Glueball}.
\newblock {\em Phys. Rev. Lett.}, 111(9):091601, 2013.

\bibitem{Gui:2019dtm}
Long-Cheng Gui, Jia-Mei Dong, Ying Chen, and Yi-Bo Yang.
\newblock {Study of the pseudoscalar glueball in $J/\psi$ radiative decays}.
\newblock {\em Phys. Rev. D}, 100(5):054511, 2019.

\bibitem{Dudek:2006ej}
Jozef~J. Dudek, Robert~G. Edwards, and David~G. Richards.
\newblock {Radiative transitions in charmonium from lattice QCD}.
\newblock {\em Phys. Rev. D}, 73:074507, 2006.

\bibitem{Yamanaka:2021xqh}
Nodoka Yamanaka, Atsushi Nakamura, and Masayuki Wakayama.
\newblock {Interglueball potential in lattice SU(N) gauge theories}.
\newblock {\em PoS}, LATTICE2021:447, 2022.

\bibitem{Yamanaka:2019yek}
Nodoka Yamanaka, Hideaki Iida, Atsushi Nakamura, and Masayuki Wakayama.
\newblock {Glueball scattering cross section in lattice SU(2) Yang-Mills theory}.
\newblock {\em Phys. Rev. D}, 102(5):054507, 2020.

\bibitem{Klempt:2007cp}
Eberhard Klempt and Alexander Zaitsev.
\newblock {Glueballs, Hybrids, Multiquarks. Experimental facts versus QCD inspired concepts}.
\newblock {\em Phys. Rept.}, 454:1--202, 2007.

\bibitem{Crede:2008vw}
V.~Crede and C.~A. Meyer.
\newblock {The Experimental Status of Glueballs}.
\newblock {\em Prog. Part. Nucl. Phys.}, 63:74--116, 2009.

\bibitem{Chen:2022asf}
Hua-Xing Chen, Wei Chen, Xiang Liu, Yan-Rui Liu, and Shi-Lin Zhu.
\newblock {An updated review of the new hadron states}.
\newblock 4 2022.

\bibitem{CrystalBarrel:1992qav}
E.~Aker et~al.
\newblock {The Crystal Barrel spectrometer at LEAR}.
\newblock {\em Nucl. Instrum. Meth. A}, 321:69--108, 1992.

\bibitem{Parganlija:2013xsa}
Denis Parganlija.
\newblock {Mesons, PANDA and the scalar glueball}.
\newblock {\em J. Phys. Conf. Ser.}, 503:012010, 2014.

\bibitem{Amsler:1995tu}
Claude Amsler and Frank~E. Close.
\newblock {Evidence for a scalar glueball}.
\newblock {\em Phys. Lett. B}, 353:385--390, 1995.

\bibitem{Sarantsev:2021ein}
A.~V. Sarantsev, I.~Denisenko, U.~Thoma, and E.~Klempt.
\newblock {Scalar isoscalar mesons and the scalar glueball from radiative $J/\psi$ decays}.
\newblock {\em Phys. Lett. B}, 816:136227, 2021.

\bibitem{Klempt:2021wpg}
Eberhard Klempt and Andrey~V. Sarantsev.
\newblock {Singlet-octet-glueball mixing of scalar mesons}.
\newblock {\em Phys. Lett. B}, 826:136906, 2022.

\bibitem{10.1093/ptep/ptaa104}
Particle~Data Group, P~A Zyla, et~al.
\newblock {Review of Particle Physics}.
\newblock {\em Progress of Theoretical and Experimental Physics}, 2020(8), 08 2020.
\newblock 083C01.

\bibitem{Bugg:2004xu}
D.~V. Bugg.
\newblock {Four sorts of meson}.
\newblock {\em Phys. Rept.}, 397:257--358, 2004.

\end{thebibliography}

    \end{document}